\documentclass[doublecol]{epl2} 

\usepackage{epsfig}
\usepackage{graphicx} 
\usepackage{dcolumn}  
\usepackage{bm}       
\usepackage{amssymb}

\title{Breaking chirality in nonequilibrium systems on the lattice}
\shorttitle{Europhys.~Lett., {\bf 81} (2008) 10009} 

\author{Diego Paz\'o\inst{1} \and Ernesto M. Nicola\inst{2}}
\shortauthor{Europhys.~Lett., {\bf 81} (2008) 10009}

\institute{                    
  \inst{1} Instituto de F\'{\i}sica de Cantabria, IFCA (CSIC-UC) - 
Avda.~Los Castros, 39005 Santander, Spain\\
  \inst{2} Max-Planck-Institut f\"ur Physik komplexer Systeme - 
N\"othnitzer Stra{\ss}e 38, 01187 Dresden, Germany
}

\abstract{We study the dynamics of fronts in parametrically forced oscillating lattices.
Using as a prototypical example the discrete Ginzburg-Landau equation, we show that much information about front bifurcations can be extracted by projecting onto a cylindrical phase space.
Starting from a normal form that describes the nonequilibrium Ising-Bloch bifurcation in the continuum and using symmetry arguments, we derive a simple dynamical system that captures the dynamics of fronts in the lattice.
We can expect our approach to be extended to other pattern-forming problems on lattices.}

\begin{document}

\maketitle

Extended systems on lattices have played a major role 
in the development of nonlinear science. 
One may recall celebrated models as the discrete sine-Gordon equation 
or the Fermi-Pasta-Ulam experiment.
Usually the development of the field has been associated with
conservative systems, but dissipative lattices 
have attracted growing attention in the last two decades (see \cite{Scott} for a review).
Two prominent examples of such lattices are provided by the discrete version of the Nagumo and Ginzburg-Landau partial differential equations.
The former has been proposed as a model of
myelination of neuronal fibres~\cite{Keener};
and the latter describes, among others, 
dissipative solitons~\cite{ds},
the dynamics of lines of vortices~\cite{willaime91b}
and coupled wakes \cite{legal91} in hydrodynamics. 

The complex Ginzburg-Landau equation~\cite{kramer}
universally describes the dynamics of an extended medium in the 
neighbourhood of an oscillatory instability. 
Under homogeneous resonant $n:1$ forcing, different regions of space 
may lock to the driving with different phase relations; 
and domain walls separate these regions. 
The prototypical $2:1$ resonant case leads to the so called parametrically forced complex Ginzburg-Landau (FCGL) equation.
This equation has been subject of extensive study since the seminal work by Coullet and coworkers
in ref.~\cite{Coullet90}.
There it was found a front bifurcation which is the nonequilibrium analogue of the Ising-Bloch transition in ferromagnets.

In this Letter, we show that the dynamics of fronts in the
FCGL equation {\em on the lattice}
is captured by a normal form consisting 
of two ordinary differential equations. 
The bifurcations linking different dynamics of the front (including bistable regimes) 
are observed both in a projection of the system's variables onto a cylindrical phase space and in the normal form.
Our results are relevant for 
experiments where discretisation is given as in arrays of coupled pendula \cite{denardo92,huang96},
electronic circuits \cite{bode,vicente}, or chemical systems \cite{laplante}; and also
in systems usually modeled as continuous,
but that are intrinsically discrete (or behave like a lattice 
due to a spatially periodic modulation of the medium).

For a lattice, the FCGL equation \cite{Walgraef,MS06}
takes the form:
\begin{eqnarray}
\dot A_j &=&(1+i\nu)A_j-(1+i\beta)|A_j|^2 A_j+\gamma A_j^*\nonumber\\
& &+ \kappa (1+i\alpha)(A_{j+1}+A_{j-1}-2A_j) \,,
\label{Eq:FCGLE_discretized}
\end{eqnarray}
where $A_j\equiv \rho_j e^{i \psi_j}$ is a complex variable.
The parameter $\gamma$ measures the forcing strength, and $\kappa$ 
controls the coupling between neighboring units.
Parameters $\nu$, $\beta$, and $\alpha$ account for the detuning, 
the nonisochronicity, and the dispersion, respectively.

\begin{figure*}
\onefigure[width=0.9\textwidth]{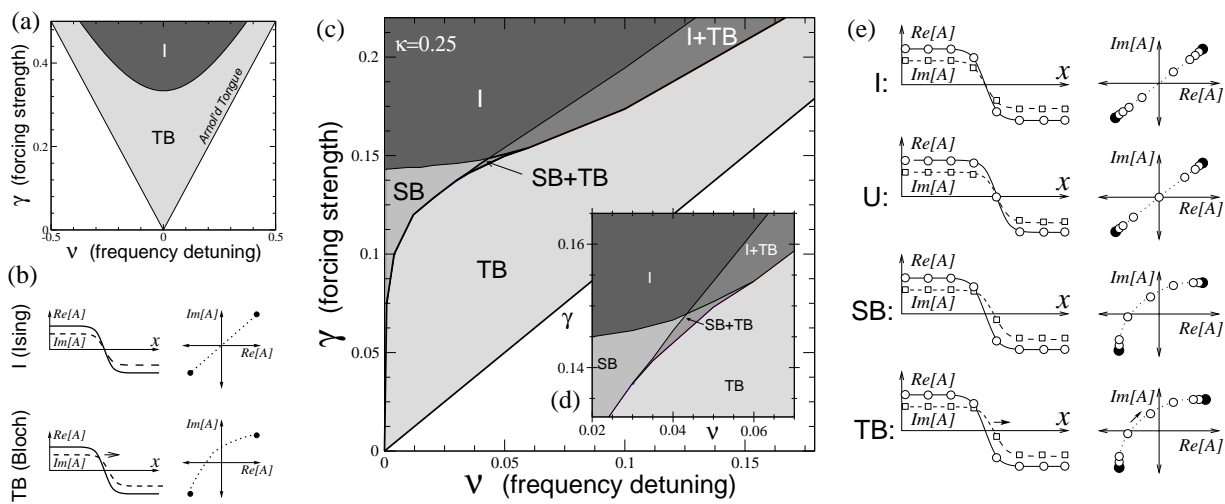}
\caption{Regions in the parameter space $\nu$-$\gamma$ where 
Ising (I) and travelling or stationary Bloch fronts 
(denoted by TB and SB, respectively) appear:  (a) in the continuum,
and (c) on the lattice, $\kappa=0.25$ in eq.~(\ref{Eq:FCGLE_discretized}).
Sketches of all possible fronts (b) in the continuum, and (e) on the lattice.
Whereas the fronts I, TB and SB in (e) are stable for some parameter values, the front 
U is always unstable.}
\label{Fig:Arnold_Tongue}
\end{figure*}

\section{The continuum limit}
First of all we recall the results for the 
continuous version of (\ref{Eq:FCGLE_discretized}):
\begin{equation}
\partial_t A=(1+i\nu)A-(1+i\beta)|A|^2 A+\gamma A^*+(1+i\alpha)\Delta A \,.
\label{Eq:FCGLE}
\end{equation}
Vanishing values of $\nu$, $\beta$, and $\alpha$
allow to cast (\ref{Eq:FCGLE}) into a variational form:
$\partial A / \partial t = - \delta \mathcal{F} / \delta A^*$.
In this case, stable front solutions minimise the energy functional $\mathcal{F}$,
and depending on the forcing $\gamma$ 
they can be chiral ($\gamma< \gamma_{IB}=1/3$) or achiral ($\gamma \ge \gamma_{IB}$).
If $A$ vanishes at the centre of the front the so-called {\em Ising} front is found,
otherwise the front is chiral and $A$ does not vanish anywhere: 
two such {\em Bloch} solutions with opposite chirality exist.

The nonequilibrium Ising-Bloch (NIB) transition is observed for nonzero values of $\nu$, $\beta$, 
or $\alpha$ ~\cite{Coullet90,tirapegui96}.
The terms $\nu$, $\beta$ and $\alpha$ are perturbations to a gradient system,
and cause the Bloch fronts to move, in either direction 
depending on their chirality (see \cite{MS06} and references therein). 

\section{Front dynamics}
We restrict ourselves hereafter to the case $\beta=\alpha=0$, so that
$\nu$ remains as the parameter that breaks the variational character of the system; 
we note nevertheless that the same qualitative results 
are obtained perturbing variationality through
$\nu$, $\beta$, and $\alpha$.
Inside the region $\gamma>|\nu|$ (the \emph{Arnold tongue}), 
see fig.~\ref{Fig:Arnold_Tongue}(a), the local dynamics
is bistable with 
two stable fixed points
$A_{S\pm} =\pm\rho_S e^{i \psi_S}$ with
$\rho_S= [1+\sqrt{\gamma^2-\nu^2}]^{1/2}$,
and $\psi_S \in (-\pi/2,\pi/2)$ the solution of 
$\sin (2 \psi_S) =\nu/\gamma$,
$\cos (2 \psi_S) = (\rho_S^2 - 1)/\gamma$. 
The fixed point at the origin $A_O=(0,0)$
is either completely unstable ($\gamma^2  < 1+ \nu^2$), or saddle ($\gamma^2  > 1+ \nu^2$). 
We are interested in the dynamics of fronts connecting the two 
stable fixed points: $A_{j\rightarrow\mp\infty}=A_{S\pm}$.

In the continuum, the locus of the NIB transition 
can be calculated analytically
~\cite{Skryabin01}: 
$\gamma_{NIB}(\nu)=[\sqrt{1+9\nu^2}]/3$, see fig.~\ref{Fig:Arnold_Tongue}(a) [and fig.~\ref{Fig:Arnold_Tongue}(b) for a sketch of both front types]. 
As mentioned above, for $\nu=0$ we recover the variational case, $\gamma_{NIB}(\nu=0)=\gamma_{IB}$, 
and Bloch fronts are stationary.

Our extensive numerical simulations of the FCGL equation on the lattice\footnote{The 
numerical calculations were performed with
zero flow boundary conditions $A_0=A_1$, $A_{N+1}=A_N$, and
a lattice size $N$ large enough to neglect boundary effects on the 
front dynamics (with $N$ typically being 128).}
have revealed several front types not present in the continuum. 
These fronts are sketched in fig. 1(e) and figs. 1(c,d)
show their regions of stability.
Stationary Bloch fronts (SB) exist on a finite region of 
parameter space around an interval of the line $\nu=0$.
Additionally, two regions of bistability are found: 
In one of them Ising and travelling Bloch fronts coexist (I+TB). 
The other bistable region (SB+TB) is shown in fig.~1(d), and 
it is a small triangle 
with stable stationary and travelling Bloch fronts.

\section{Cylindrical coordinates}
A projection of the $2N$ degrees of freedom onto a two-dimensional phase space 
greatly simplifies the analysis of the fronts.
One way of performing this projection is
\begin{eqnarray}
\Phi &=&  \mathrm{Re}[  \sum_{j=1}^N \frac{\rho_j}{\rho_S} e^{i(\psi_j-\psi_S)}  ] \,, \label{Eq:cp1}\\
C &=&     \mathrm{Im}[  \sum_{j=1}^N \frac{\rho_j}{\rho_S} e^{i(\psi_j-\psi_S)}  ] \,. \label{Eq:cp2}
\end{eqnarray}
This corresponds to a rotation and compression
 of the complex plane $A$ such that the stable 
fixed points are now located on the real axis at $\pm1$.
This projection permits us to discern if a front is symmetric with respect to the origin
(in such a case $C=0$). Note that
$\Phi$ is a cyclic variable that takes the same value when the front
advances or recedes one lattice unit (provided the front is far from the
boundaries) and thus can be defined modulo $2$.
Variable $C$ measures the deviation
from stationarity and is intrinsically bounded.

Typical examples of the dynamics in the reduced coordinates (\ref{Eq:cp1})-(\ref{Eq:cp2}) are shown in figs.~2(a,b).
Two stationary fronts, both with $C=0$, exist for all parameter values.
One, the Ising front (I), is located at 
$(\Phi,C)=(0,0)$ if the number of units $N$ is even (or at $(1,0)$ if $N$ is odd)
and is a continuation for $\kappa > 0$ of the 
trivial solution at zero coupling:  
$[\ldots,A_{S+}, A_{S+},A_{S+},A_{S-},A_{S-},A_{S-},\ldots ]$.
The second stationary solution, denoted U, is
at $(1,0)$ (respectively, at $(0,0)$ for odd $N$), and 
is a continuation of the unstable front 
solution: $[\ldots,A_{S+}, A_{S+},A_O,A_{S-},A_{S-},\ldots ]$.
For some parameter values, there exist extra fixed points 
located off the $\Phi$ axis. 
We label these chiral solutions stationary Bloch (SB) fronts.
Travelling Bloch (TB) fronts correspond to 
periodic orbits around the cylinder. 
Due to symmetry they always appear in pairs circulating in opposite directions.

\begin{figure}
\onefigure[width=0.45\textwidth]{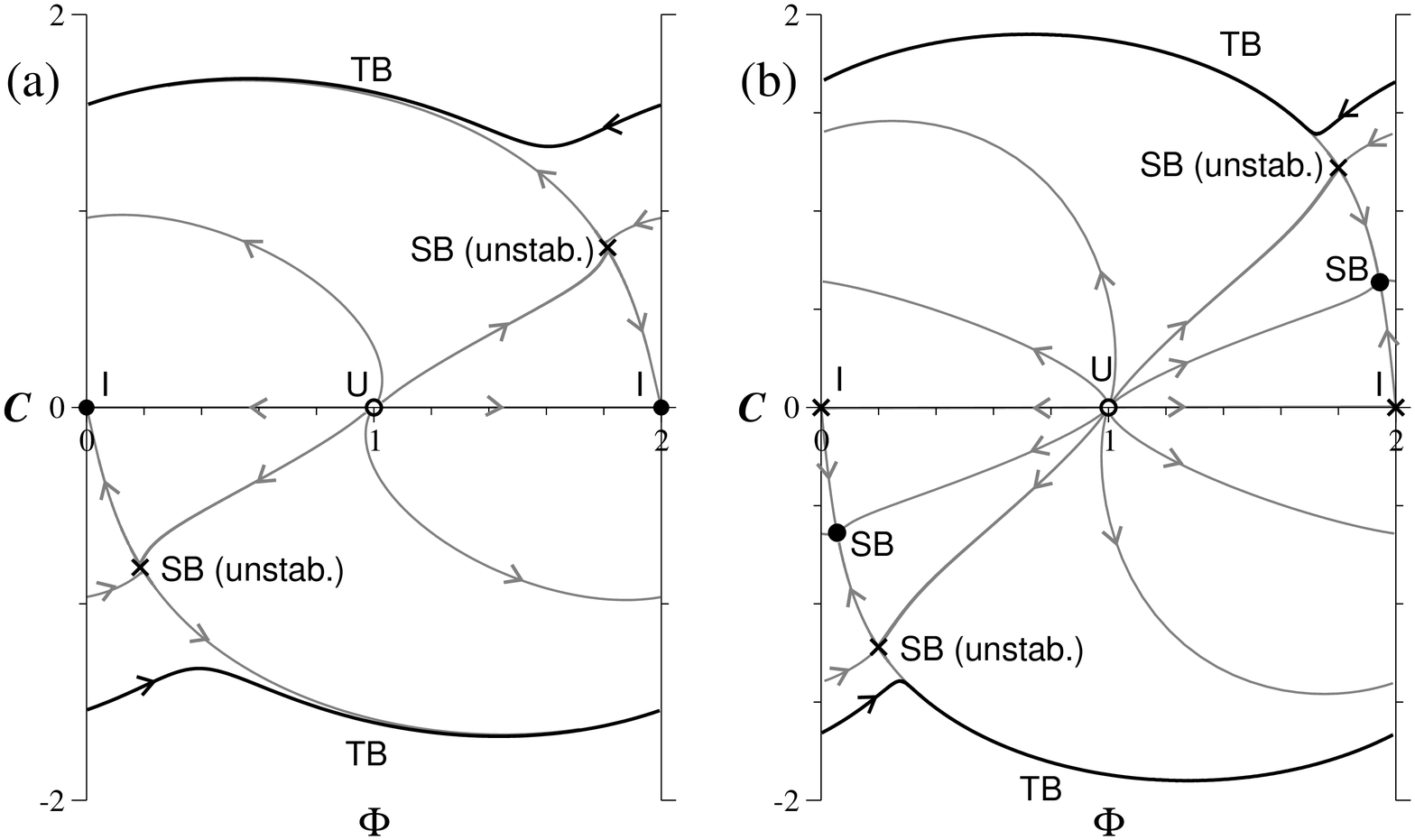}
\caption{Numerically obtained flows projecting the front dynamics 
onto the cylindrical phase space $\Phi$-$C$ defined by
eqs.~(\ref{Eq:cp1}) and (\ref{Eq:cp2}).
In (a) an example of coexistence {I+TB} is shown ($\nu=0.1$ and $\gamma=0.183$).
In (b) {SB+TB} ($\nu=0.045$ and $\gamma=0.148$).
Symbols $\bullet, \times, \circ$ denote stable, saddle and 
unstable fixed points, respectively. }
\label{Fig:sketch}
\end{figure}

\section{Normal form}
Next, we present a simple ordinary-differential-equation model 
that generates dynamics like the front dynamics on the FCGL lattice.
We will obtain this normal form  via symmetry arguments. 

In continuous systems the NIB transition is a parity-breaking (pitchfork) bifurcation 
coupled to a translation-invariant coordinate. 
It is an example 
of the so-called drift-pitchfork bifurcation
found in a number of situations 
({\it e.g.}~\cite{gollub83}),
usually as a secondary instability. 
Its normal form is \cite{tb_o2,kness} 
$\dot \phi=c$, $\dot c = (\mu-c^2)c$, 
where $\mu$ is the bifurcation parameter, e.g., $\mu \propto (\gamma_{NIB}-\gamma)$.
Coordinates $\phi$, $c$ represent the position 
and the velocity of the concerned structure (the front in our case). 

Knobloch {\it et al.}~\cite{knobloch95} considered the breakdown
of the {\em continuous} translational invariance
to study the parity breaking of a periodic pattern in the presence 
of an inhomogeneity.
This leads to the inclusion of small periodic terms (sinusoidals in the simplest
case) that preserve the invariance under inversion $(\phi,c)\rightarrow(-\phi,-c)$:
\begin{eqnarray}
\dot \phi &=& c- \epsilon \sin\phi \,, \label{nfa}\\
\dot c &=& (\mu+\delta \cos\phi -c^2)c +\eta \sin\phi \,,
\label{nfb}
\end{eqnarray}
where $\phi$ is now an angular variable. 
In~\cite{pazo_pla} it was suggested that this normal form could also be used to analyse 
parity-breaking bifurcations found in discrete bistable media ({\it e.g.} arrays of 
FitzHugh-Nagumo units).

In the continuum, as the variational limit is approached
the velocity of the Bloch front decreases (and becomes zero at variationality).
Consequently, we introduce a small parameter $\chi$ that 
accounts for the deviation from the variational case.
For the situation considered here, $\chi$
is proportional to\footnote{According to Eq.~(4) in \cite{Coullet90}, in the variational limit one has $\chi \propto \nu -\beta -(\alpha-\beta)~\gamma$.} $\nu$. 
We have then at leading order:
\begin{eqnarray}
\dot \phi &=& \chi c- \epsilon \sin\phi \,, \label{nf2a}\\
\dot c &=& (\mu+\delta \cos\phi -c^2)c \label{nf2b} \,.
\end{eqnarray}
The term $\eta \sin\phi$  present in (\ref{nfb}) is  absent in (\ref{nf2b})
due to the invariance of the discrete FCGL equation
under the transformation $(\nu,\beta,\alpha,A) \rightarrow (-\nu,-\beta,-\alpha,A^*)$.
This requires the symmetry under $(\chi,\phi,c) \rightarrow (-\chi,\phi,-c)$ 
to be satisfied.
This considerably simplifies the normal form.

\begin{figure}
\onefigure[width=0.4\textwidth]{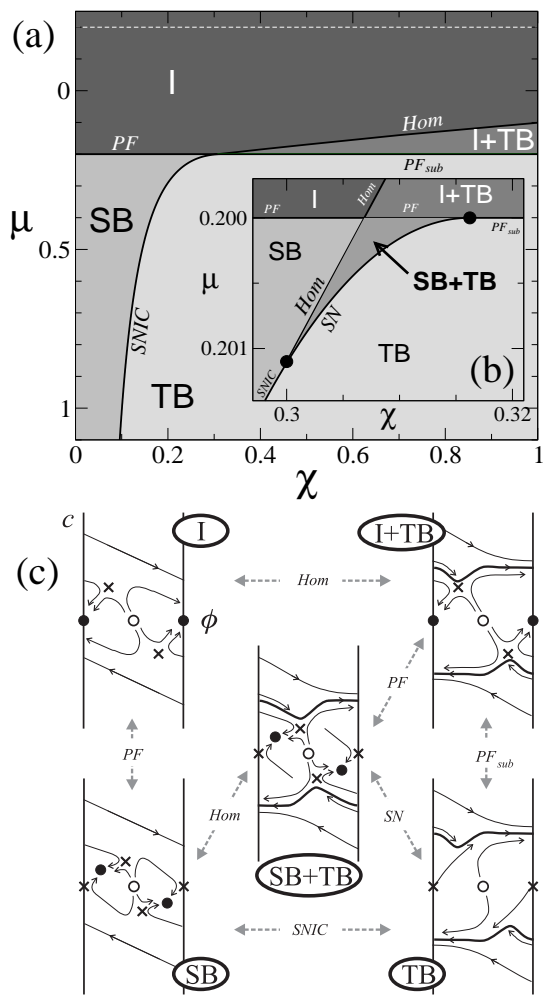}
\caption{(a,b) Parameter space of the normal form (\ref{nf2a})-(\ref{nf2b})
with $\epsilon=0.1$ and $\delta=-0.2$; the loci of local and global bifurcations are
depicted with solid lines.
The black dots in the inset (b) indicate the location of two codimension-2 bifurcation points. 
(c) Schematic representation of the phase space $\phi$-$c$
for the five regions in (a,b). 
Gray-dashed lines indicate the link between different states via 
codimension-1 bifurcations ({\it PF} and $PF_{sub}$ are, respectively, super- and subcritical pitchfork, 
and {\it SN}, {\it Hom}, and {\it SNIC} stand for (off-cycle) saddle-node, homoclinic, and 
saddle-node on an invariant circle bifurcations).}
\label{Fig:sketch_nf}
\end{figure}

Figure \ref{Fig:sketch_nf}(a) shows the regions of the parameter space $\chi$-$\mu$,
and fig.~\ref{Fig:sketch_nf}(c) is a sketch of 
the corresponding phase spaces. 
The normal form (\ref{nf2a})-(\ref{nf2b})
qualitatively reproduces front dynamics, which 
can indeed be projected on a cylinder, 
{\it e.g.}~the phase space $\Phi$-$C$ introduced above. 
There are two pitchfork bifurcations at $\mu=\pm\delta$ and the
parameter space is organised by two codimension-2 points. We may see in fig.~\ref{Fig:sketch_nf}(b)
that at a degenerate pitchfork point located at $\hat \chi^2= -2\epsilon^2/\delta $ 
the pitchfork bifurcation line switches between subcritical and supercritical. 
The second codimension-2 point is a
saddle-node separatrix-loop point 
\cite{schecter87} where the saddle-node bifurcation on the invariant circle ({\it SNIC})
splits into an off-cycle saddle-node ({\it SN}) and a 
homoclinic ({\it Hom}) bifurcations.

In contrast to the continuous case, travelling Bloch fronts appear with nonzero chirality, 
and the (average) velocity of the front $v$ asymptotically
follows: (i) the familiar square-root law when crossing the {\it SNIC} line; 
(ii) a logarithmic law
below the homoclinic line:
$v^{-1}=a - (1/\lambda_u) \ln (\gamma-\gamma_{Hom})$, 
where $\lambda_u$ is 
the positive eigenvalue corresponding to the saddle Bloch front.

Let us finally discuss the parameters modelling the discretisation strength: $\epsilon$ and $\delta$.
In the continuum limit, our numerical simulations indicate that 
discretisation enters in the normal form through $\delta$
at order $O(\kappa^{-1})$, whereas $\epsilon$ vanishes much faster
(possibly exponentially) as $\kappa \rightarrow \infty$.
In addition, assuming opposite signs for $\epsilon$ and $\delta$ 
correctly inherits the bifurcations in the anticontinuum limit ($\kappa \rightarrow 0$).
Accordingly, our choice $\epsilon=0.1$ and $\delta=-0.2$ in fig.~\ref{Fig:sketch_nf}(a) satisfies 
these constraints
({\it i.e.}~$|\epsilon| < |\delta|$, $\epsilon\delta<0$). 
The organisation of the parameter space is qualitatively robust 
to changes of both parameters in a wide range.

\section{Phase equation}
Normal form (\ref{nf2a})-(\ref{nf2b}) is obtained
in a perturbative way including small terms that model 
breakdown of translational invariance, and
the 
departure from $(\nu,\beta,\alpha;\gamma)=(\mathbf{0};\gamma_{IB})$, the equilibrium Ising-Bloch bifurcation point. 
This means that we cannot predict the behaviour
of the bifurcation lines far from this point. In particular, the ({\it SNIC\/}) bifurcation line, which separates
SB and TB regions in fig.~\ref{Fig:Arnold_Tongue}(c), can be proved to end at $\nu=\gamma=0$ as
results from the following facts. For small $\gamma$,
a phase reduction~\cite{Coullet90,MS06} of the FCGL lattice yields a discrete Nagumo-type equation
[the overdamped Frenkel-Kontorova (FK) model]:
$\dot \psi_j= \nu - \gamma \sin (2\psi_j) + \kappa (\psi_{j+1}+ \psi_{j-1} -2\psi_{j} )$.
Interestingly, $\nu$ acts as a symmetry-breaking parameter, but the symmetry
of the original model is hidden in the factor 2 inside the $\sin (2\psi_j)$ term, 
which allows two mirror front
solutions connecting $\psi_{j\rightarrow -\infty}=\psi_S$ 
and $\psi_{j\rightarrow \infty}=\psi_S+\pi$. 
Discreteness implies the existence of an interval of `propagation failure'~\cite{Keener}
where propagation is blocked for nonvanishing $\nu$. 
Well established results for the FK model (see e.g.~\cite{kladko00,carpio03}) state 
that the propagation threshold ({\it SNIC} line) should approach 
the $\gamma$ axis exponentially fast\footnote{Recent results for 
a different nonlinearity suggest however that the threshold vanishes at 
particular values (`pinning failure') what would imply that the {\it SNIC} line touches the
$\gamma$ axis at some values, and possibly a singular behaviour in the $\gamma \rightarrow 0$
limit; see~\cite{elmer06} for details.} as $\gamma \rightarrow 0$.

\section{Conclusions}
The nonequilibrium Ising-Bloch bifurcation in the parametrically driven 
complex Ginzburg-Landau equation is one of the most studied 
pattern instabilities. In the continuum, breaking of chirality 
causes the front to move. However, as shown here, 
on the lattice a more complex scenario appears: specifically, 
two types of bistability
and a region with chiral (Bloch) stationary fronts. We have demonstrated
that a normal form with two variables captures the dynamics of the front
and the bifurcations between different regions.
It is to be emphasised that the normal form (\ref{nf2a})-(\ref{nf2b})
is based on symmetry arguments that provide general qualitative results 
independent of details as, for instance,
the parameter of nonvariationality or the discretisation-order of
the Laplacian.

The dynamics of patterns on lattices is typically much harder to solve analytically than in their continuous counterparts.
The continuum, usually serves as a zeroth order approximation that is not necessarily accurate.
Discreteness typically introduces new dynamical regimes 
as shown in the current letter for one spatial dimension\footnote{
In two dimensions an even higher number of dynamical regimes can be expected 
to appear due to discreteness, since the continuous FCGL equation already exhibits a variety of patterns such as spirals, labyrinths, stripes, hexagons, etc (see {\it e.g.} \cite{Coullet1992}). 
Discreteness should give rise to a plethora of exotic patterns, such as spiral waves with extended defects \cite{pre_izus}.}.
Through a modification of the normal form for the continuum, we have been able to reproduce 
the dynamics of fronts on a discrete medium and the structure of parameter space.
This approach should also work in other problems on lattices.

\acknowledgments

We thank Manuel A.~Mat{\'{\i}}as, Juan M.~L\'opez, Ernest Montbri\'o and Luis G.~Morelli 
for useful discussions and critical comments.
D.P.~acknowledges support by {\em Ministerio de Educaci\'on y Cultura} (Spain) 
through the Juan de la Cierva Programme. 
This work was supported by MEC (Spain) 
under Grant No.~FIS2006-12253-C06-04.

\newcommand{\noopsort}[1]{} \newcommand{\printfirst}[2]{#1}
  \newcommand{\singleletter}[1]{#1} \newcommand{\switchargs}[2]{#2#1}


\end{document}